\documentclass[preprint,showpacs,preprintnumbers,amsmath,amssymb]{revtex4}

\usepackage{graphicx}% Include figure files
\usepackage{dcolumn}% Align table columns on decimal point
\usepackage{bm}% bold math

\begin{document}

%\preprint{APS/123-QED}

\title{ Superconductivity in $MgB_2$ doped with $Ti$ and $C$}

\author{R. H. T. Wilke$^{\dagger\ddagger}$}
\author{S. L. Bud'ko$^{\dagger\ddagger}$}%
\author{P. C. Canfield$^{\dagger\ddagger}$}%
\author{M. J. Kramer$^{\ddagger}$}%
\author{Y. Q. Wu$^{\ddagger}$}%
\author{D. K. Finnemore$^{\dagger\ddagger}$}%
\email{finnemor@ameslab.gov} \affiliation{
$^{\dagger}$Ames Laboratory and $^{\ddagger}$Department of Physics and Astronomy\\
Iowa State University, Ames, IA 50011}
\author{R. J. Suplinskas and J. V. Marzik}
\affiliation{Specialty Materials, Inc., 1449 Middlesex Street,
Lowell, MA 01851}
\author{S. T. Hannahs}
\affiliation{National High Magnetic Field Laboratory, Florida State University\\
1800 E. Paul Dirac Drive, Tallahassee, FL 32310}

\date{\today}

\begin{abstract}

Measurements of the superconducting upper critical field,
$H_{c2}$, and critical current density, $J_c $, have been carried
 out for $MgB_2$ doped with $Ti$ and/or $C$ in order to explore the
problems encountered if these dopants are used to enhance the
superconducting performance. Carbon replaces boron on the $MgB_2$
lattice and apparently shortens the electronic mean free path of
$MgB_2$ and raising $H_{c2}$. Titanium forms precipitates of
either $TiB$ or $TiB_2$ that enhance the flux pinning and raise
$J_c$. Most of these precipitates are intra-granular in the
$MgB_2$ phase.  For samples containing both $C$ and $Ti$ doping,
the $C$ appears to still replace $B$ in the $MgB_2$ lattice and
the $Ti$ precipitates out as a boride.  If approximately $0.5\%
Ti$ and approximately $2\% C$ are co-deposited with B to form
doped boron fibers and these fibers are in turn reacted in $Mg$
vapor to form doped $MgB_2$, the resulting superconductor has $\mu
_oH_{c2}(T=0)~\sim ~25~T$ and $J_c~\sim ~10,000A/cm^2$ at $5~K$
and $2.2~T$.

\end{abstract}

% PACS, the Physics and Astronomy Classification Scheme.
\pacs{74.25.Bt, 74.25.Fy, 74.25.Ha}

%Use showkeys class option if keyword display desired
\keywords{carbon doping, titanium doping, magnesium diboride,
superconductivity}

\maketitle

\section{Introduction}

The performance of $MgB_2$ superconducting materials\cite {1} can
be greatly enhanced by the addition of small amounts of carbon
that will raise the upper critical magnetic field, $H_{c2}$,\cite
{2} and the critical current density, $J_c$.\cite {3}  Point
defects like carbon that substitute for boron in the host lattice
have been shown to raise $\mu _oH_{c2}(T=0)$ from $16~T$ in pure
$MgB_2$ to $32~T$ for a carbon level of $\sim 4\%$.\cite {4}  At
higher carbon contents, $\mu _oH_{c2}$ rises more slowly and
eventually drops to about $25~T$ at $\sim 10\%$ carbon.\cite {5,
6}  Several different precipitate phases have been used to form
small pinning sites in the $MgB_2$ lattice.  For example,
$SiC$,\cite {7} $YB_4$,\cite {8}, $TiB_2$\cite {9} and
$MgSi_2$\cite {10} have been added and show a rise of $J_c$ into
the range of $10^6 ~A/cm^2$ at $10~K$ and self field with about
$5\%$ precipitate additions.  In many of these experiments, either
a powder-in-tube $(PIT)$ process or a pressed and sintered pellet
method have been used to form the precipitates for the synthesis.

In a different approach to sample preparation, a chemical vapor
deposition, $CVD$, method\cite {11} can be used to co-deposit $B$
together with a doping element to form long lengths of carbon
doped boron fiber.\cite {4} Subsequent heat treatment in $Mg$
vapor transforms the doped boron into doped $MgB_2$.\cite {4}
Both the $CVD$ and powder methods have advantages.   The advantage
of powder methods is that the diffusion lengths are comparable to
the powder size giving relatively low reaction temperatures and
short reaction times. The advantage of co-depositing the impurity
with the $B$ in a $CVD$ process is that the impurity is mixed with
the $B$ on an atomic scale.  Both $Ti$\cite {3} and $C$\cite {4}
separately have been successfully doped into $MgB_2$ using these
$CVD$ methods.

The purpose of the work reported here is to study the combined $C$
and $Ti$ doping of $MgB_2$ to determine whether samples can be
prepared with the combined benefits of both dopants, raising $\mu
_oH_{c2}$ to the range of $30~T$ and raising $J_c$ values to the
range of $10^4~A/cm^2$ at $20~K$ and $5~T$. Two specific questions
need to be addressed. Does the addition of $Ti$ to the $C$-doped
$MgB_2$ samples reported earlier,\cite {4} reduce the amount of
$C$ in the $MgB_2$ lattice, possibly by forming $TiC$
precipitates?  In the presence of $C$, can the $Ti$ precipitate
size be maintained at $\sim 5~nm$ for high $J_c$ performance?

\section{Experimental}

Long lengths of doped boron fiber are prepared in a $CVD$
apparatus similar to that used for commercial boron filament
production.\cite {11}  All of the materials reported here were
deposited on a $W$ fiber with an initial diameter of about $15\mu
m$. The $W$ enters the reaction chamber moving at a few $cm/s$
through a $Hg$ seal into a long glass tube containing flowing
$H_2$, $BCl_3$, $CH_4$, and $TiCl_4$.  For $Ti$ doping, $H_2$ is
bubbled through liquid $TiCl_4$ to provide $TiCl_4$ molecules in
the gas stream.  Typically the $BCl_3$ flow rate was $3~l/min$
with the $CH_4$ and $TiCl_4$ flow rates adjusted for the desired
doping levels. The diameter of the exiting doped $B$ fibers were
in the $75\mu m$ to $100\mu m$ range.  The doped $B$ fibers were
cut to lengths of $\sim 3~cm$ and placed in a $Ta$ tube with a
$Mg$ to $B$ ratio of $1:1$. The $Ta$ was welded shut, sealed in a
quartz tube, and heat treated in a box furnace for appropriate
periods. Upon removal from the furnace, ampoules were quenched in
water. Resistivity measurements are made by a four contact method
using silver epoxy to make the electrical contacts. A Quantum
Design $PPMS$ system was used to make resistivity vs temperature,
$\rho ~vs.~T$, measurements up to $14~T$.  At higher fields, $\rho
~vs.~H$ measurements are made in a $32.5~T$ copper coil magnet at
the National High Magnetic Field Laboratory at Florida State
University.  Magnetization measurements were made in a Quantum
Design SQUID magnetometer with a magnetic field range of $5.5~T$
and $J_c$ was determined from the magnetization hysteresis using
the Bean Model for cylindrical shell samples.\cite {3}  A Philips
CM30 transmission electron microscope ($TEM$) was employed for the
microstructure characterization. Samples for the $TEM$  were made
using a crush-flow technique.

\section {$Ti$-doping only}

In an earlier publication,\cite {12} we investigated a series of
$Ti$-doped $MgB_2$ samples that were deposited on a commercial
carbon coated $SiC$ fiber substrate with a diameter of
approximately $80~\mu m$ instead of the $15~\mu m$ diameter $W$
substrate used here. The doped-$MgB_2$ layers ranged from about
$4~\mu m$ to $10~\mu m$ thick. Flux pinning was excellent and gave
$J_c$ over $10,000~A/cm^2$ at $25~K$ and $1.3~T$ for a sample with
an average doping of about $9\% Ti$.\cite {3} Unfortunately, there
were problems with this approach to sample preparation.  First,
the diffusion of $Mg$ into the $B$ caused swelling and the $MgB_2$
pulled away from the substrate. Second, the presence of a carbon
coated substrate made the amount of $C$ in the specimen uncertain.
In addition, these samples had a rather inhomogeneous $Ti$
distribution. There was a high $Ti$ level near the $SiC$ core and
near the outer surface of the fiber.\cite {3} Transmission
electron microscope $(TEM)$ observations for one of these samples
that had been reacted for $2~h$ at $950^{\circ }C$ are shown in
Fig. 1a. Results reveal a random distribution of precipitates
scattered through the grain, ranging in size from $1~nm$ to
$20~nm$ with a spacing about 5 times as large. Energy dispersive
spectroscopy for a large portion of the grain in Fig. 1a, revealed
that the $Ti/Mg$ ratio was $\sim 5\% $. Selected area diffraction
taken along the c-axis showed the prominent hexagonal pattern of
$MgB_2$ with additional powder pattern rings arising from the
titanium boride precipitates. The beam was then tipped off axis to
show the rings more prominently. Indexing the rings revealed that
the precipitates were $TiB$ rather than $TiB_2$. These results
differ from those found by Zhao and coworkers.\cite {13} They
found that samples prepared by mixing powders of $Mg$, $Ti$, and
$B$ and reacting gave $TiB_2$ precipitates on the $MgB_2$ grain
boundaries. This illustrates that different sample preparation
methods can yield different types and locations of the
precipitates.  Both the precipitation of $TiB$ throughout the
grains\cite {12} and precipitation of $TiB_2$ on the grain
boundaries\cite {13} seem to give enhanced flux pinning.

Samples with $15~\mu m$ diameter tungsten cores were prepared
using three different $TiCl_4$ flow rates in the $CVD$ reactor,
$0.42~cc/min$, $1.26~cc/min$ and $2.9~cc/min$. After the $CVD$
deposition, the fibers were reacted in $Mg$ vapor to form
$Ti$-doped $MgB_2$. Energy dispersive spectra (EDS) in a scanning
electron microscope (SEM) was used to probe the uniformity of $Ti$
distribution across the superconducting fiber and to measure the
$Ti$ to $Mg$ ratio, $[Ti]/[Ti+Mg]$.  Line scans in various regions
of the sample show that the $Ti$ level was uniform to about $10\%
$ of the average value.  Multiple point scans and area scans in
the EDS give average values of the $[Ti]/[Ti+Mg]$ ratio, shown in
Fig. 2 indicating that the percent $Ti$ is roughly linearly
related to the flow rate in the reaction chamber. Point scans,
line scans, and large area raster scans were used for analysis.
The large area raster scan shown by the open squares on Fig. 2
comprises the most data and are probably the most accurate measure
of the $Ti$ content. These three flow rates of $0.42~cc/min$,
$1.26~cc/min$, and $2.9~cc/min$, give $0.3\% $, $0.5\% $ and
$1.6\% $ respectively for the $[Ti]/[Ti+Mg]$ ratio.

Because these $Ti$ doped samples were deposited on a $15~\mu m$
diameter $W$ wire and had a $76~\mu m$ outer diameter, these
fibers require much longer times or higher temperatures to fully
form the $MgB_2$ phase than the $4$ to $10~\mu m$ $B$ layers of
the earlier work.\cite {3} Therefore, temperatures of $1000$ to
$1200^{\circ }C$ were often used. A sample with $0.5\%Ti$ reacted
for $72~h$ at $1000^\circ C$ were found to be $95\%$ reacted using
polarized light in an optical microscope. A $TEM$ micrograph of
this sample, Fig. 1b, shows precipitates that are much larger than
in Fig. 1a. In this figure, the sweeping shaded areas arise from
the underlying holey carbon support. In this micrograph, the beam
has been tilted to emphasize areas of dense dislocations as shown,
for example, in the top center of the micrograph. Many of the
precipitates were $50$ to $200~nm$ in diameter. Selected area
diffraction indicated the precipitates were $TiB_2$ and had c-axes
coaxial with the $MgB_2$ grains in which they were imbedded. For a
sample with $0.5\%Ti$ plus $2.1\%C$, the sample was fully reacted
after $48~h$ at $1200^{\circ }C$ and shows the $20$ to $100~nm$
diameter precipitates of Fig. 1c. Again the large precipitates
here are $TiB_2$ with the c-axis parallel to the c-axis of the
$MgB_2$ grain in which it is imbedded.

Another series of samples having $0.3\% $, $0.5\% $ and $1.6\%~Ti$
were all reacted at $1100^{\circ }C$ for $48~h$ and studied in the
$TEM$.  For $0.3\%Ti$, the precipitate size is less than $50~nm$
and widely spaced.  For $0.5\%Ti$, the precipitate size ranges
from $20$ to $80~nm$.  For $1.6\%Ti$, the precipitate size ranges
from $20$ to $100~nm$ and the density of precipitates is
correspondingly higher.  All of these samples show $J_c$ values
higher than pure $B$, but lower than shown by the sample of Fig.
1a.

Magnetization data, shown in Fig. 3, indicate that the suppression
of $T_c$ with $Ti$ doping depends on the temperature at which the
$MgB_2$ forms.  Samples reacted at $1000^{\circ }C$ for $72~h$
have magnetization curves similar to pure $MgB_2$ except that they
are shifted to lower temperature by about $1~K$. The $0.3\% Ti$
and $1.6\% Ti$ are nearly identical whereas the $0.5\% Ti$ sample
is a bit lower.  The depression is not monotonic in $Ti$ content,
and the results on Fig. 3 probably represent true scatter in the
data.  The cause of the suppression in $T_c$ is not understood. If
the $Ti$ all precipitates as $TiB$ or $TiB_2$ and no $Ti$ is
incorporated in the $MgB_2$ lattice, then it might be expected
that the suppression of $T_c$ might be rather small and arise from
lattice strains induced by the precipitates or other defects in
the $MgB_2$.  It is also possible that some $Ti$ atoms replace
$Mg$ in the $MgB_2$ lattice and reduce the superconducting
interaction in that way. Further study is going to be needed to
determine which variables contribute to this suppression in $T_c$.

At $1100^{\circ }C$ for $48~h$, the suppression of $T_c$ behavior
in Fig. 3b is similar to Fig. 3a except that the downward shift of
$T_c$ is somewhat larger. At $1200^oC$ for $24~h$, shown in Fig.
3c, the suppression of $T_c$ is very large and increases
monotonically with $Ti$ content. Because the $4\pi M~vs.~T$ curves
are not always monotonic in the $Ti$ content, a new series of
samples were made at $1100^{\circ }C$ to check reproducibility.
The data for the two sets of samples with the same $Ti$ content
differed by as much as $0.5~K$.  With these large $TiB_2$
precipitates, there are some as yet uncontrolled parameters.

An X-ray study of the shift in the $MgB_2$ a-axis lattice constant
was undertaken to look for a connection between the lattice
constant an the amount of impurity.  For the case of carbon,\cite
{4} the a-axis lattice constant contracts linearly with increasing
$C$, as shown by Fig. 4, while the c-axis remains essentially
unchanged as reported by several authors.\cite {4} For the case of
$Ti$, the picture is more complicated.  As shown in Fig. 4, the
a-axis contracts with increasing $Ti$, but there is considerable
scatter in the data and the amount of change in the a-axis may
depend on the temperature at which the $MgB_2$ is formed.  An
additional sample with $4.7~cc/min$ flow rate of $TiCl_4$
($2.4\%Ti$) and reacted at $1200^{\circ }C$ for $12~h$ is shown by
the open square of Fig. 4. The apparent flat region of the a-axis
lattice parameter between $0.5\%$ and $2.4\%~Ti$ as shown by the
open squares would appear to indicate that a solubility limit has
been reached for $Ti$ in $MgB_2$ at about $0.5\%~Ti$.

For the case of $Ti$ additions only, samples with an $1100^{\circ
} C$ for $48~h$ reaction temperature are used to illustrate the
changes in superconducting properties. The values of $H_{c2}$
measured up to $9~T$ for the three different $Ti$ concentrations
are slightly lower than for pure $MgB_2$ as shown on Fig. 5. As
was found for values of $4\pi M~vs.~T$ curves in Fig. 3b, the
$0.5\%~Ti$ sample has the largest suppression of $T_c$ and the
largest suppression of $H_{c2}$.

As shown by Fig. 6, the $J_c$ values at $20~K$ for these three
$Ti$ concentrations are enhanced by about a factor of $10$ at
fields, up to about $1.5~T$.  There is very little difference in
$J_c$ as the $Ti$ level is raised from $0.3\% Ti$ to about $1.6\%
Ti$. Values of $J_c$ drop through $1~kA/cm^2$ at about $1.5~T$
even though $\mu _oH_{c2}$ is above $6~T$ at $20~K$ for these
three samples as shown in Fig. 5.

\section {Combined $Ti$ and $C$ doping}

All the samples reported here for the combined doping were reacted
at $1200^{\circ }C$ to ensure full conversion to the $MgB_2$
phase.  The $C$ content is determined from the $CH_4$ flow rate in
the $CVD$ chamber and previous results.\cite {4}  For a sample
with $0.5\%Ti$ plus $2.1\%C$ shown in Fig. 1c, the precipitates
are $20$ to $100~nm$ in diameter and selected area diffraction
shows the precipitates to be $TiB_2$ with the c-axis parallel to
the c-axis of the $MgB_2$ grain in which they are imbedded.  In
this micrograph, the beam is tilted so that the dislocations are
not so apparent, but they are there.  These precipitates of Fig.
1c are similar to those in Fig. 1b, but in contrast to the $TiB$
precipitates of Fig. 1a.

An X-ray analysis of the combined $Ti$ and $C$ doping is
illustrated in Fig. 7. The c-axis lattice parameter from the (002)
peak essentially does not change with $Ti$ and $C$ addition.  The
a-axis lattice parameter contracts in a regular way as shown by
the (110) peak of Fig. 7.  The $0.5\%~Ti$-only peak moves to
higher angle than the pure $MgB_2$ peak by about $0.11$ degree as
shown by the heavy dotted line.  The $2.1\%~C$-only peak shifts
out from the pure $MgB_2$ peak by about $0.22$ degree as shown by
the light dotted line. And, the $0.5\%~Ti$+$2.1\%~C$ peak shifts
out from the pure $MgB_2$ peak by about $0.29$ degree as shown by
the heavy solid line.  Roughly speaking, the decrease in a-axis
lattice parameter is additive for $Ti$ and $C$ doping at this
level.

As was reported earlier,\cite {4} the $4\pi M~vs.~T$ curves and
the resulting $T_c$ values are depressed monotonically with
increasing $C$ content as shown by the open squares and solid
circles of Fig. 8. If an additional $0.5\%~Ti$ is added to each of
these $C$ concentrations, the combined suppression of $T_c$ is
roughly additive.  As shown in Fig. 8, the addition of $2.1\% ~C$
to $MgB_2$ suppressed $T_c$ by about $2~K$ and the addition of
$2.1\% ~C$ plus $0.5\% Ti$ suppressed $T_c$ by about $6~K$.

Values of $\mu _oH_{c2}$ in samples with the combined doping shown
in Fig. 9 by the open circles are very similar to values for $C$
doping only reported previously\cite {4} as shown by the solid
symbols. The solid triangles were taken with $R~vs.~T$
measurements and the solid circles were taken at the National High
Magnetic Field Laboratory as $R~vs.~H$ measurements.  For the
combined $Ti$ and $C$ doped samples there was considerable
rounding at the high field end of the $R~vs.~H$ transitions.  The
two open circles represent two different definitions of $\mu
_oH_{c2}$, the lower being a linear extrapolation of the long
linear region of $R~vs.~H$ up to the normal state, and the upper
open circle being the field where the resistivity reaches the
normal state value within the noise.  The addition of a few
percent $C$ to the $Ti$ doped samples substantially raises $\mu
_oH_{c2}$ (open circles) to values comparable to  values for
carbon only (solid symbols).

Values of $J_c$ for a series of samples reacted at $1200^{\circ
}C$ are shown in Fig. 10.  For the combined $2.1\%~C$ + $0.5\%~Ti$
sample at $5~K$, the $J_c$ curve crosses $1~kA/cm^2$ at about
$3.2~T$ shown by the solid circles.  The $1.1\%~C$ + $0.5\% ~Ti$
sample at $5~K$ crosses  $1~kA/cm^2$ at $2.6~T$, as shown by the
open circles.  The $0.5\%Ti$ only sample has the highest low field
$J_c$ values and crosses $1~kA/cm^2$ at $2.3~T$.  A pure $MgB_2$
sample at measured at $5~K$ is shown by the solid squares.  It is
not shown here, but the the combined $2.1\%~C$ + $0.5\%~Ti$ sample
at $20~K$, the $J_c$ curve crosses $1~kA/cm^2$ at about $1.5~T$.

To summarize, the addition of $C$ to raise $H_{c2}$ and the
addition of $Ti$ to form precipitate pinning centers are roughly
independent of one another for the samples reported here.   Carbon
doped $MgB_2$ shows a rapid rise in $\mu _oH_{c2}(T=0)$ from
$16~T$ for pure $MgB_2$ to $25~T$ for $\sim ~2.1\%$ added carbon.
With both $2.1\%$ $C$ and $0.5\%~ Ti$ the sample retains a $\mu
_oH_{c2}(T=0)$ of $\sim 25~T$ and no evidence for the formation of
$TiC$ was seen.  The addition of $Ti$ enhances $J_c$ substantially
in the magnetic field range of $2.5$ to $3.5~T$, as shown in Fig.
10.

Much work needs to be done to optimize the the $C$ and $Ti$ levels
and the processing to raise $J_c$ at high fields. For low reaction
temperatures and short times to form the $MgB_2$ phase, $TEM$ data
show that the precipitates are intra-granular and randomly
oriented particles ranging in size from $1$ to $20~nm$. For higher
reaction temperatures in the range from $1000$ to $1200^{\circ
}C$, the precipitates are much larger.  The precipitates are
usually intra-granular $TiB_2$ particles coplanar with the $MgB_2$
host that range from $50~nm$ to $200~nm$ in size.  Coarsening of
the $TiB_2$ precipitates at high temperatures and long times is
clearly a problem.  In magnetic fields from zero to $1~T$, the
addition of carbon to $Ti$-doped $MgB_2$ gives relatively little
change in $J_c$, but in the $3$ to $4~T$ range, carbon additions
clearly enhance $J_c$.  A very practical problem is that the
addition of either $C$ or $Ti$ slows down the rate at which the
$B$ fibers transform to the $MgB_2$ phase. For high $J_c$ values,
it is helpful to react at low temperatures to give small $MgB_2$
grains and to prevent the $Ti$ precipitate coarsening. Some method
is needed to overcome these slow reaction rates, probably the use
of fine powders to keep the reaction time short and the reaction
temperatures low. It would be desirable to make doped boron
powders so that the diffusion lengths can be much smaller and the
reaction temperatures lower.

\begin{acknowledgments}
Ames Laboratory is operated for the US Department of Energy by
Iowa State University under Contract No. W-7405-Eng-82. This work
was supported by the Director for Energy Research, Office of Basic
Energy Sciences. A portion of this work was performed at the
National High Magnetic Field Laboratory, which is supported by NSF
Cooperative Agreement No. DMR-0084173 and by the State of Florida.
\end{acknowledgments}

\eject

\clearpage

FIGURE CAPTIONS

Fig. 1  $TEM$ micrograph of $MgB_2$ with $5\% Ti$ reacted
$950^{\circ }C$-$2~h$, $C$ coated $SiC$ substrate. b) $MgB_2$ with
$0.5\% Ti$ reacted $1000^{\circ }C$-$72~h$, $W$ substrate. c)
$MgB_2$ with $0.5\%Ti+2.1\%C$ reacted $1200^{\circ }C$-$48~h$, $W$
substrate.

Fig. 2  Percent $Ti$ in $MgB_2$ sample as a function of the
$TiCl_4$ flow rate in the $CVD$ apparatus.

Fig. 3  Magnetization transitions for 3 different $Ti$ levels at
a) $1000^oC$, b) $1100^{\circ }C$ and $1200^{\circ }C$.

Fig. 4  Comparison of change in a-axis lattice constant for both
$Ti $ and $C$ doping.

Fig. 5  $H_{c2}$ for $MgB_2$ doped with $Ti$ only.

Fig. 6  Enhancement of $J_c$ with $Ti$ additions.

Fig. 7  X-ray data for the (002) and (110) peaks for pure B,
$0.5\%~Ti$, $2.1\%C$, and $0.5\%Ti+2.1\%C$.

Fig. 8  Comparison of $C$ + $Ti$ doping with $C$ only.

Fig. 9  $H_{c2}$ for combined $C$ + $Ti$ doping (open symbols)
compared to $C$ only doping (solid symbols).

Fig. 10  Enhancement of $J_c$ for combined $Ti$ and $C$ additions.
The $Ti$ only curve was reacted $1200^{\circ }C~12h$.  The $Ti$
plus $C$ samples were reacted at $1200^{\circ }C~48h$.

\clearpage

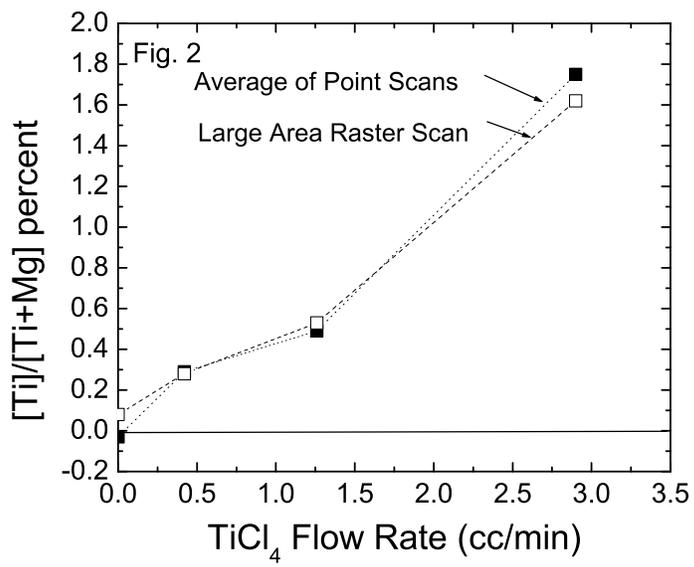
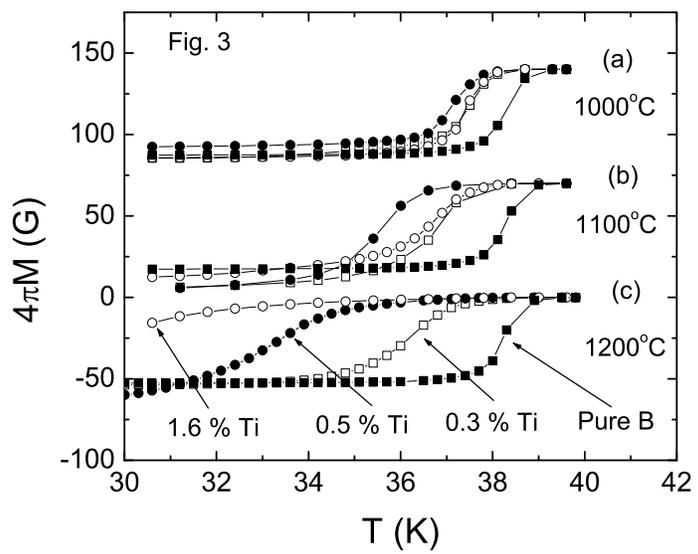
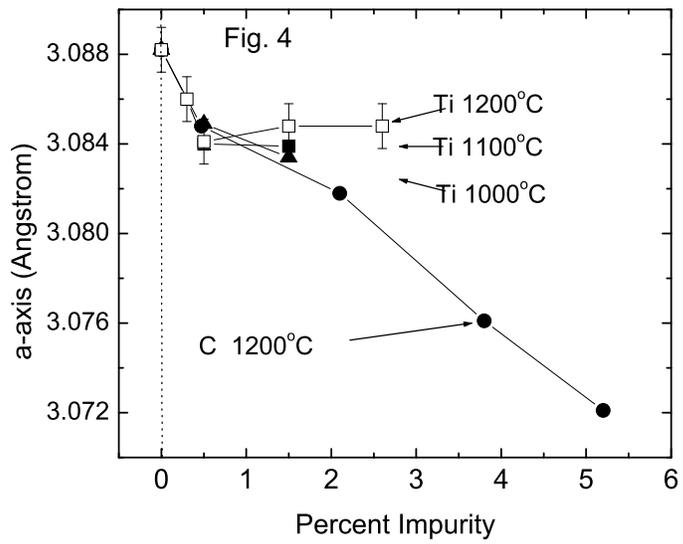
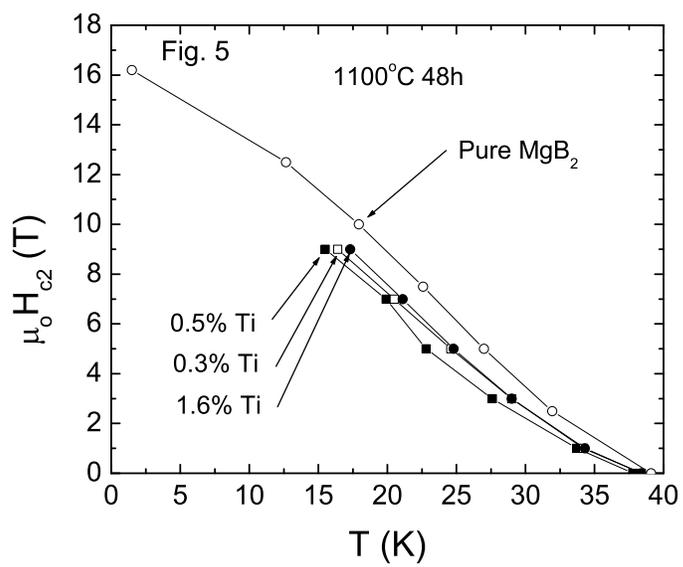
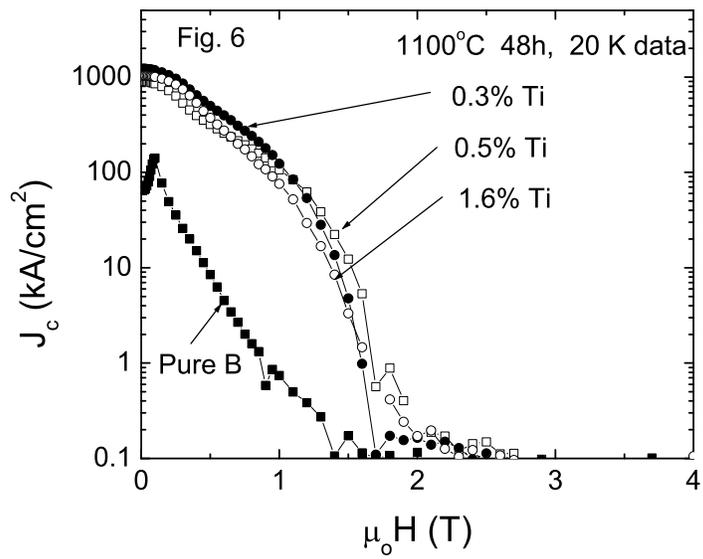
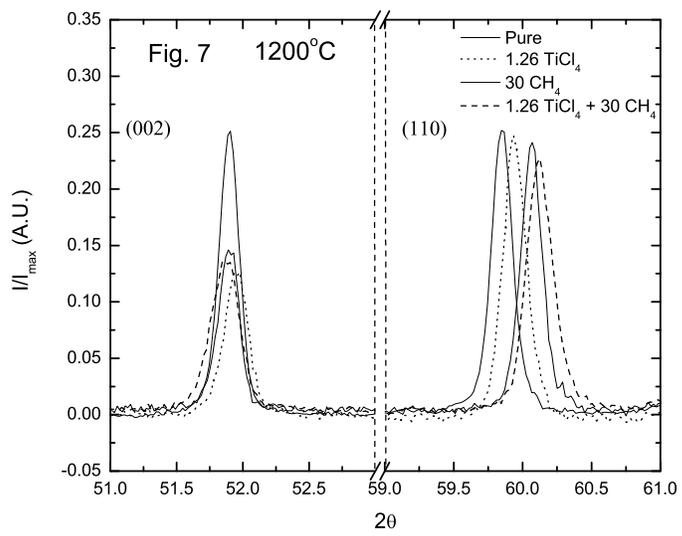
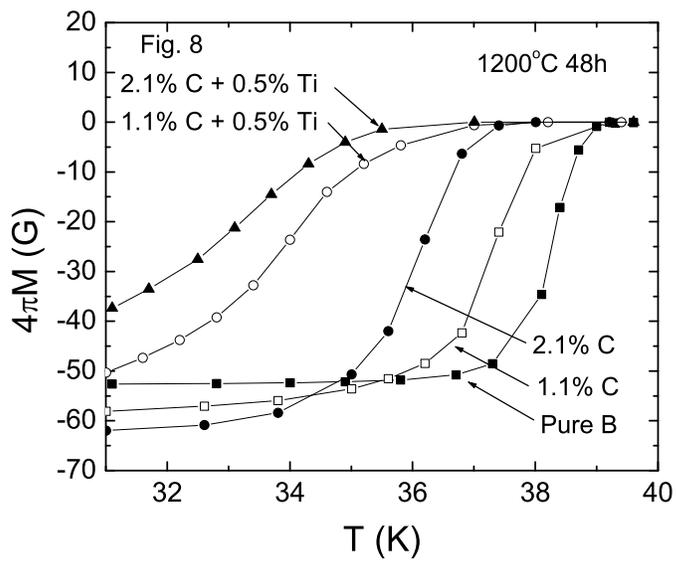
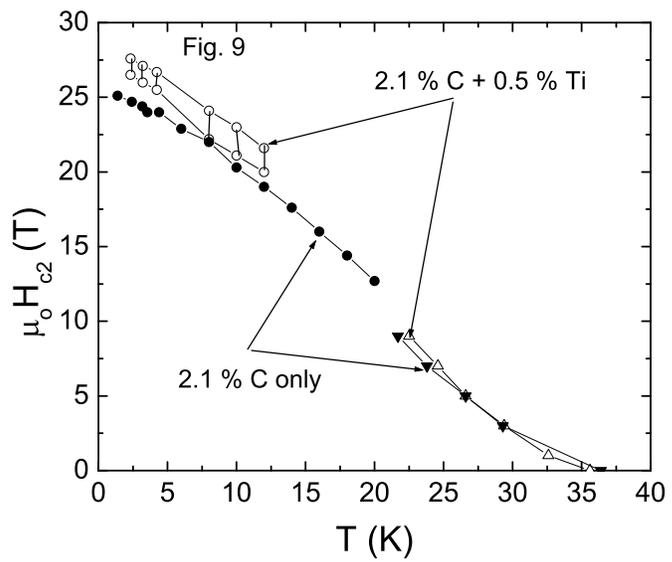
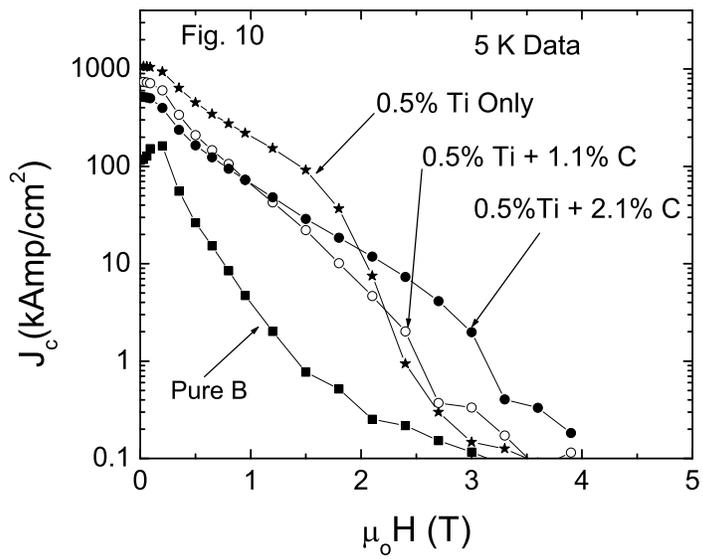
\end{document}